\documentclass{iopconfser}

\begin{document}

\title{Leveraging the null stream to detect strongly lensed gravitational wave signals}

\author{Jef Heynen$^{1}\orcidlink{0009-0007-5496-2159}$, Soumen Roy$^{1,2}\orcidlink{0000-0003-2147-5411}$ and Justin Janquart$^{1,2}\orcidlink{0000-0003-2888-7152}$}

\affil{$^1$Centre for Cosmology, Particle Physics and Phenomenology - CP3, Université Catholique de Louvain, Louvain-La-Neuve, B-1348, Belgium \newline
}
\affil{$^2$Royal Observatory of Belgium, Avenue Circulaire, 3, 1180 Uccle, Belgium}

\email{jef.heynen@uclouvain.be}

\begin{abstract}
Gravitational lensing of gravitational waves is expected to be observed in current and future detectors. In view of the growing number of detections, computationally light pipelines are needed. Detection pipelines used in past LIGO-Virgo-KAGRA searches for strong lensing require parameter estimation to be performed on the gravitational wave signal or are machine learning based. Removing the need for parameter estimation in classical methods would alleviate the ever growing demand of computational resources in strong lensing searches and would make real-time analysis possible. We present a novel way of identifying strongly lensed gravitational wave signals, based on the null stream of a detector network. We lay out the basis for this detection method and show preliminary results confirming the validity of the formalism. We also discuss the next development steps, including how to make it independent of parameter estimation.
\end{abstract}

\section{Introduction}
Like electromagnetic radiation, gravitational waves (GWs) can be gravitationally lensed by massive objects along their trajectory \cite{Schneider:1992}. In general, lensing will produce a frequency-dependent modulation to the signal \cite{Takahashi:2003}. However, when the GW wavelength is much smaller then the Schwarzschild radius of the lens, lensing leaves the frequency evolution of GW unaffected, corresponding to the \textit{geometric optics} regime. In this case, multiple images of the signal are produced, but they are magnified and time- and phase-shifted with respect to the original signal \cite{Takahashi:2003,Dai:2017}. When the time difference between the images is large enough such that they do not overlap in the sensitive frequency band of the detector, we speak of \textit{strong lensing}. Searches for such repeated signals have not yielded any compelling evidence so far \cite{Hannuksela:2019,LIGOScientific:2023,Janquart:2023}. These searches, conducted by the LIGO-Virgo-KAGRA collaboration, comprise methods based on parameter estimation (PE) of the signals or on machine learning (see \cite{ Wright:2025} for an overview of the methods). Here, we present a novel way of searching for signals of strongly lensed GWs, based on the \textit{null stream} of a detector network.

In Section \ref{sec: strong lensing} and Section \ref{sec: null stream}, we lay out the basics of strongly lensed GWs and the null stream formalism. Since the frequency evolution of strongly lensed images are identical, one can construct a joint null stream, if one accounts properly for the magnfication, time shift and Morse phase shift. We explain how this can be done in Section \ref{sec: nullstream-lensing}. Finally, in Section \ref{sec: results}, we show the validity of this idea by means of an injection study, and explain the perspectives for this method.

\section{Strong lensing}
\label{sec: strong lensing}
As discussed above, for strongly lensed GWs, the geometric optics approximation applies and multiple images of the same signal will arrive on Earth \cite{Takahashi:2003}. Since the apparent sky separation between different images is unresolvable with GW detectors, the GW signals have matching parameters, apart from those parameters biased by strong lensing. This is because strong lensing, although it leaves the frequency evolution of the lensed images unaffected, causes separation in time $\delta t_j$ and will have a (de)magnification $\mu_j$ for the $j$-th image with respect to the original signal. There is a third effect, which depends on the type of extremum of the time delay function that the image corresponds to. This effect introduces a three-valued quantity $n_j=\{0,\frac{1}{2},1\}$, called the \textit{Morse factor}, corresponding to a minimum, saddle point, or maximum. The lensed waveform in frequency domain of the $j$-th image, $\tilde{h}^j_L(f)$, will take the form \cite{Dai:2017}
\begin{equation}
    \tilde{h}^j_L(f)=\sqrt{\mu_j}\,e^{-2\pi i f \delta t_j+i\pi n_j \text{sign}(f)}\tilde{h}_{U}(f)\,.
\end{equation}
where $\tilde{h}_U(f)$ is the waveform of the GW in the absence of a lens.

\section{Null stream of a detector network}
\label{sec: null stream}
If one knows the sky location of the GW source exactly, one can linearly combine the data of $D > P$ detectors such that any signal component is cancelled in the result, where $P$ is the number of polarizations in the theory. The result is referred to as a null stream \cite{Chatterji:2006,Sutton:2009}.
Assuming a true sky position $\hat{\mathbf{\Omega}}$ of the GW source, the data in detector $\alpha=1,\dots, D$ is
\begin{equation}
    \label{eq: data in freq domain as sum of two polas and noise}
    \tilde{d}_\alpha(f)=e^{-2\pi i f \Delta t_\alpha(\hat{\mathbf{\Omega}})}\left[F_\alpha^+(\hat{\mathbf{\Omega}},\psi)\tilde{h}_+(f) + F_\alpha^\times(\hat{\mathbf{\Omega}},\psi) \tilde{h}_\alpha^\times (f)\right] +\tilde{n}_\alpha(f)\,.
\end{equation}
where $n_\alpha$ is the noise in detector $\alpha$, which we assume to be Gaussian and stationary, $F_\alpha^+\,,F_\alpha^\times$ are the antenna response functions for detector $\alpha$ and are functions of the sky position $\hat{\mathbf{\Omega}}$ and the polarization angle $\psi$, and $\Delta t_\alpha(\hat{\mathbf{\Omega}})$ is the time delay of the signal in detector $\alpha$ at position $\vec{r}_\alpha$ with respect to some reference position $\vec{r}_0$. Henceforth, we work with whitened noise $\tilde{n}_{\mathrm{w}\alpha}(f)$ and the noise weighted quantities $\tilde{d}_{\mathrm{w}\alpha}(f)$ and $F_{\mathrm{w}\alpha}^{+,\times}(\hat{\mathbf{\Omega}},\psi,f)$, which means we divide by the square root of the one-sided power spectral density $S_\alpha(f)$.
We can now rewrite Eq. \eqref{eq: data in freq domain as sum of two polas and noise} in terms of the noise weighted vectors $\tboldd$, $\boldF$, $\tboldh$, and $\tboldn$ as
\begin{equation}
    \label{eq: data as sum of polas and noise vector form}
    \tboldd = \boldF\tboldh + \tboldn\,,
\end{equation}
as done in \cite{Sutton:2009}.

A projector onto the \textit{signal space}, spanned by the antenna response matrix, and a projector onto its orthogonal, the \textit{null space}, are given by
\begin{equation}
    \Pgw = \bm{F}\left(\bm{F}^\dagger \bm{F}\right)^{-1}\bm{F}^\dagger\,,\quad \Pnull = 1 - \Pgw\,,
\end{equation}
respectively. The null projector allows us to remove the signal from the data, thereby leaving only noise, which can be statistically characterized, without the need of knowing the exact signal shape.
Lastly, we define the null energy for a generic sky position $\mathbf{\Omega}$ as the magnitude of the null stream summed over each frequency bin:
\begin{equation}
    \label{eq: null energy for generic sky position}
    E_\text{null}(\mathbf{\Omega}) = \sum_{f} \tboldd^\dagger(f)\Pnull(\mathbf{\Omega}, f) \tboldd(f)\,.
\end{equation}
When $\mathbf{\Omega}$ matches the true sky position $\hat{\mathbf{\Omega}}$, this reduces to 
\begin{equation}
    \label{eq: chi squared}
    E_\text{null}(\hat{\mathbf{\Omega}}) = \sum_{f} \tboldn^\dagger(f)\Pnull(\hat{\mathbf{\Omega}}, f) \tboldn(f) \sim \chi_{(D-2)N}\,,
\end{equation}
where the chi-squared distribution holds for Gaussian and stationary noise and $N$ is the number of frequency bins. Notice that when we replace the data stream by the signal $\bm{F}\tboldh$, the null energy will be zero for the true sky position.

\section{Null stream based strong lensing detection statistic}
\label{sec: nullstream-lensing}
Suppose we have $M$ GW events, where event $j$ is observed with $D_j$ detectors, corresponding to lensed images of the same source. This means they are related via the lensing parameters, and Eq. \eqref{eq: data as sum of polas and noise vector form} can be adapted as follows. We stack the data and noise streams of different images in one vector and define the antenna response matrix as
\begingroup
\renewcommand*{\arraystretch}{1.5}
\begin{align}
    \label{eq: F-matrix for strong lensing}
    \boldF = \begin{pmatrix}
    F_1^{(1)+},\dots,F_{D_1}^{(1)+},F_{1}^{(2)+},\dots,F_{D_2}^{(2)+},\dots,F_{1}^{(M)+},\dots,F_{D_M}^{(M)+}\\
    F_1^{(1)\times},\dots,F_{D_1}^{(1)\times},F_{1}^{(2)\times},\dots,F_{D_2}^{(2)\times},\dots,F_{1}^{(M)\times},\dots,F_{D_M}^{(M)\times}
    \end{pmatrix}^T\,,
\end{align}
\endgroup
with
\begin{align}
    \begin{split}
        F_\alpha^{(1)+,\times} \equiv F_\alpha^{(1)+,\times}(t_1, \mathbf{\Omega}, \psi)\,,\quad
        F_\alpha^{(j)+,\times} \equiv \sqrt{\mu_{j1}} e^{i\pi n_{j1} \text{sign}(f)}  F_\alpha^{+,\times(j)}(t_j, \mathbf{\Omega}, \psi),
    \end{split}
\end{align}
where $\tilde{d}_{\mathrm{\mathrm{w}}\alpha}^{(j)}$ and $\tilde{n}_{\mathrm{\mathrm{w}}1}^{(j)}$ are the whitened data and noise for image $j$ in detector $\alpha$, $\mu_{j1}=\mu_j/\mu_1$ is the relative magnification, $t_j$ the observed time of coalescence, $t_{j1}$, and $n_{j1}=n_j-n_1$ is the relative Morse factor. This equation shows that if we correct the antenna response functions for the lensing parameters of the corresponding image, only projections of the GW onto the two-dimensional subspace spanned by these corrected antenna response functions will contribute. Hence for the correct sky position and lensing parameters, $\bm{F}$ given as in Eq. \eqref{eq: F-matrix for strong lensing} can be used to construct a null stream for the joint network of $D_1+D_2+\dots+D_M$ detectors. The joint null energy can now serve as a strong lensing detection statistic: when we have a set of strongly lensed images, it is possible to construct a null stream. However, GWs that are unrelated, cannot be mapped onto each other via the lensing parameters and hence they cannot constitute a null stream.

Since the sky and lensing parameters are not known beforehand, one can use, instead of the data stream, the signal or a reconstruction thereof, to compute the null energy, divide the parameter space into a grid and look for the values which constitute the best null stream. Another approach is to use a nested sampling algorithm \cite{Skilling:2006}. We opt for the latter, providing us with posterior distributions on the relevant parameters. It also has the additional benefit that electromagnetic follow-ups can be directly performed without the need for joint PE.

\section{Preliminary results and outlook}
\label{sec: results}
To test the validity of this approach, we perform an injection study of a 1000 unrelated events and a 100 strongly lensed pairs, each event having a signal-to-noise ratio of at least 12 at LIGO and Virgo O4 design sensitivity \cite{Acernese:2015,Capote:2025}. We perform PE on all events and obtain the maximum likelihood (maxL) waveforms $\tilde{h}_L(f)$ from this. For each pair, we then perform a nested sampling procedure on the lensing and sky parameters, minimizing the joint null energy, in which we replace the data stream with the maxL waveform. This sampling procedure takes only a few minutes on one processing unit. We then take the mean of the joint null energy posterior. We repeat this to compute the mean of the null energies of the individual events, now sampling only over the sky parameters. A histogram showing the difference between the mean of the joint null energy and the individual null energies is shown in Fig. 1.
\begin{figure}[!ht]
    \centering
    \label{fig: E_null histo}
     \includegraphics[scale=0.8]{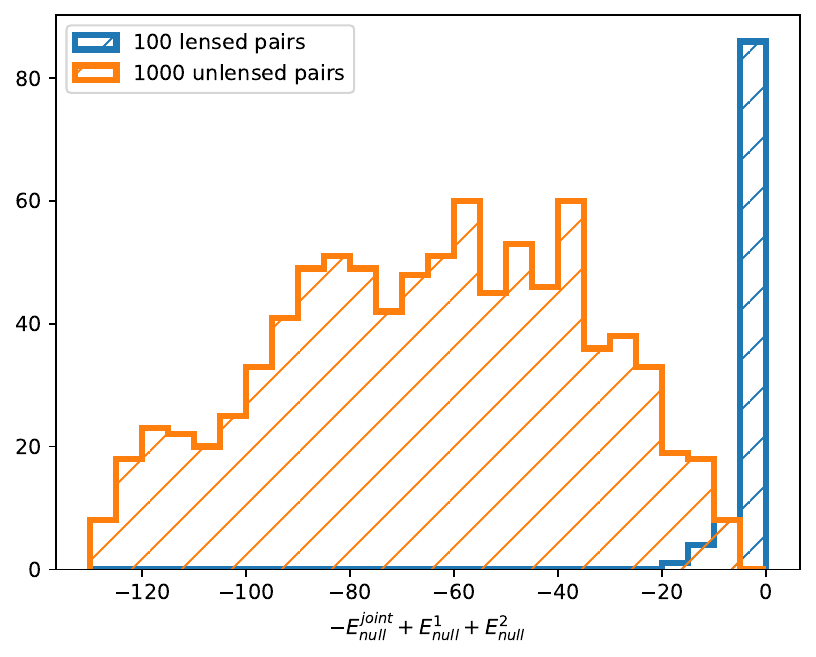}
     \caption{Comparison of the mean null energy of a 1000 unlensed event pairs with a 100 strongly lensed pairs. The bulk of the unlensed pairs have null energies deviating from zero, whereas the bulk of the lensed pairs have null energies close to zero. The overlap is due to a coincidence in waveform parameters in the unlensed case and a non perfect waveform reconstruction in the lensed case.}
\end{figure}
The plot shows the bulk of the lensed pairs to have null energy close to zero, while there are only a few unlensed pairs that have a small null energy when analyzed jointly, which is due to coincidentally similar waveforms \cite{Caliskan:2022}. There is a small overlap in the histograms, which is due to the maxL waveform not being a perfect reconstruction of the true waveform, which leads to residuals in the null energy.

We performed PE for these preliminary results in order to extract the maxL waveform. Since PE is done on supertreshold triggers regardless, we can assume such a step to be readily available to us for real events. However, we notice that a reconstructed waveform can be found without using PE, for example by a wavelet-based reconstruction \cite{klimenko:2021}. Such pipelines were originally designed for burst searches and are less accurate for longer duration signals. Preliminary investigations with \texttt{cWB} have shown this also reduces our capacity to identify strongly lensed compact binary coalescences. However, this could make such an approach favourable in strongly lensed burst searches. 

Additionally, we can make this method fully waveform independent, by working directly on the noisy data frames and using the statistical properties of the null energy, given in Eq. \eqref{eq: chi squared}, for lensed images. Because of residuals in the null stream, the null energy will not follow this distribution for unlensed events. However, for this effect to be discernable from the noise in frequency domain, loud signals are generally required. By going to the time-frequency domain, one filters out the signal-free time-frequency bins, suppressing the noise effects and optimizing the effect of residuals in the joint null stream. This method is currently being researched.
\section*{Acknowledgements}
J. H. is a FRIA grantee of the Fonds de la Recherche Scientifique - FNRS. The authors are grateful for computational resources provided by the LIGO Laboratory and supported by National Science Foundation Grants PHY-0757058 and PHY-0823459. Computational resources have been provided by the Consortium des Équipements de Calcul Intensif (CÉCI), funded by the Fonds de la Recherche Scientifique de Belgique (F.R.S.-FNRS) under Grant No. 2.5020.11 and by the Walloon Region. The authors thank Mick Wright for the careful review.
\bibliography{bibliography}

\newpage

\end{document}